\documentclass
[superscriptaddress,secnumarabic,amssymb,amsmath,nobibnotes,aps,prd,showkeys,showpacs,nofootinbib,onecolumn,12pt]{revtex4}%
\usepackage{graphicx}
\usepackage{epsf}
\usepackage{bm}
\usepackage{amsmath}
\usepackage{amsfonts}
\usepackage{amssymb}%
\providecommand{\U}[1]{\protect\rule{.1in}{.1in}}

\newcommand{\be}{\begin{equation}}
\newcommand{\ee}{\end{equation}}

\newcommand{\mincir}{\raise
-3.truept\hbox{\rlap{\hbox{$\sim$}}\raise4.truept\hbox{$<$}\ }}
\newcommand{\magcir}{\raise
-3.truept\hbox{\rlap{\hbox{$\sim$}}\raise4.truept\hbox{$>$}\ }}

\begin{document}
\title{Kantowski-Sachs cosmology in scalar-torsion theory}
\author{Andronikos Paliathanasis}
\email{anpaliat@phys.uoa.gr}
\affiliation{Institute of Systems Science, Durban University of Technology, PO Box 1334,
Durban 4000, South Africa}
\affiliation{Departamento de Matem\'{a}ticas, Universidad Cat\'{o}lica del Norte, Avda.
Angamos 0610, Casilla 1280 Antofagasta, Chile.}

\begin{abstract}
In\ the context of scalar-torsion theory we investigate the evolution of
the\ cosmological anisotropies for a Kantowski-Sachs background geometry. We
study the phase-space of the gravitational field equations by determining the
admitted stationary points and study their stability properties. For the
potential function of the non-minimally coupled scalar field we assume the
exponential and the power-law functions. Finally, we make use of Poincare
variables in order to investigate the existence of stationary points at the
infinity regime of the dynamics. 

\end{abstract}
\date{\today}
\maketitle

\section{Introduction}

\label{sec1}

In \cite{ksp1} Kantowski and Sachs, exact solutions of the field equations in
General Relativity with a spatially homogeneous, anisotropic and irrotational
spacetime were investigated. The spacetime admits a four-dimensional isometry
group transitive on three-dimensional spacelike hypersurfaces. The isometries
comprise a translation symmetry and a three-dimensional subgroup the orbits of
which are two-dimensional surfaces of constant curvature. The exact solution
determined in \cite{ksp1} is a non-vacuum solution for which the energy
momentum tensor in the Einstein field equations describes a dust fluid source.
An important characteristic of the Kantowski-Sachs spacetimes is that they can
be seen as extended Schwarzschild manifolds \cite{col1}.

The Kantowski-Sachs spacetime is related with two other line elements of
cosmological interest. Indeed, from the line element of the locally
rotationally symmetric mixmaster universe, that is, the Bianchi IX universe,
after a Lie contraction the Kantowski-Sachs element is recovered. Furthermore,
when the shear vanishes in the latter spacetime, that is, the spacetime is
isotropic, the spacetime describes the closed
Friedmann--Lema\^{\i}tre--Robertson--Walker (FLRW) space \cite{jm1}. In
\cite{cc1} it was found that the Kantowski-Sachs universe becomes isotropic
when a cosmological constant term is introduced into the field equations
\cite{cc1}, similarly to the result found by Wald in \cite{cc2}. Moreover, the
introduction of a positive cosmological constant indicates the existence of an
initial singularity for the spacetime which evolves to the de Sitter universe
\cite{cc3,cc3a}. In \cite{cc4} it was found that the cosmic no-hair conjecture
is widely valid for Kantowski-Sachs geometries which asymptotically approach
the de Sitter universe

Additionally, the dynamics of Kantowski-Sachs model can describe the
time-evolution for the physical variables in anisotropic and inhomogeneous
models \cite{sz0}. There exist a family of Szekeres solutions \cite{sz1} which
describe inhomogeneous Kantowski-Sachslike geometries without any isometry
\cite{sz2}. Kantowski-Sachs geometry can be used for the description of the
universe in the very early stages and near to the cosmological singularity.
Thus, quantum gravity in this specific anisotropic geometry has been widely
studied \cite{qq1,qq2,qq3,qq4}. There is a plethora of important results in
the literature for the Kantowski-Sachs geometry with other kinds of matter
sources, exact solution with a radiation fluid was found earlier in \cite{r1},
while a stiff fluid source was considered in \cite{r2} and later in \cite{r3}
in wh the scalar field was assumed to be that of a massless scalar field. See
also \cite{r4,r5,r6,r7,r8,r9,r10} and references therein.

The cosmological constant in the context of General Relativity is a simple
dynamical approach in order to reproduce expansion in the cosmological
parameters. Nevertheless, the cosmological constant suffers from two major
problems \cite{cca1,cca2}. Consequently, cosmologists have considered other
ways to explain the early-time and late-time acceleration phases of the
universe \cite{cca3}. There are various approaches in the literature for the
description of the observed phenomena. Modified theories of gravity form a
large family of cosmological models in which the acceleration is attributed to
geometrodynamical of freedom \cite{md2}. The latter follow from the new
invariant functions which are introduced to modify the Einstein-Hilbert Action
Integral. The modification of the gravitational Action Integral with the
introduction of the squared Ricci scalar term is a simple modification which
can provide a simple mechanism for the explanation of the inflationary epoch
\cite{star,BOtt}. This inflationary model belongs to a more general family of
theories known as $f\left(  R\right)  $-theory of gravity
\cite{Buda,Sotiriou,odin1,fr1}.

Apart from the Ricci scalar there are other invariants which have been
introduced to modify the gravitational theory \cite{mod1,mod2,mod3,mod5,mod6}.
The Ricci scalar is constructed by the Levi-Civita connection, which is the
fundamental geometric object of General Relativity. However, a more general
connection can be considered as the basis of the gravitational theory. The
Weitzenb{\"{o}}ck connection \cite{Weitzenb23} is a curvature-less connection
which leads to the Teleparallel Equivalent of General Relativity (TEGR)
\cite{cc}. In TEGR the gravitational Lagrangian is defined by the torsion
scalar $T$~\cite{Hayashi79}. On the other hand, a torsion-free connection
which describes a flat geometry and nonmetricity scalar $Q$ leads to the
so-called Symmetric Teleparallel General Relativity (STGR) \cite{mm1}. The
Kantowski-Sachs geometry has also been investigated and in modified theories
of gravity \cite{ks1,ks2,ks3,ks4,ks5,ks7}.

In this piece of study we are interested on the evolution of the cosmological
parameters in a Kantowski-Sachs background geometry in scalar-torsion theory
\cite{ss1}. Scalar-torsion is the analogue of scalar-tensor theory \cite{ks8}
in teleparallelism \cite{ks9}, in which a non-minimally coupled scalar field
is introduced into the gravitational Action Integral and it is coupled to the
torsion scalar $T$. Scalar-torsion is a theory of special interest in
cosmological studies because it provides a geometric mechanism for the
explanation of the acceleration phases of the universe \cite{ss2,ss3}, as a
unification in the dark sector components of the universe \cite{ss4}. While
there seem to be many similarities between the scalar-torsion and the
scalar-tensor theories, in terms of background dynamics \cite{ss6,ss6a}, the
two theories are complete different \cite{ss7,ss8,ss9}. Recently, in
\cite{ss10} with the use of the Noether symmetry analysis new analytic and
exact solutions for the field equations for the classical and quantum level in
scalar-torsion theory were derived for a Kantowski-Sachs geometry. In the
following, we present a detailed analysis of the evolution of the dynamics. We
study the phase-space of the cosmological parameters by determining the
stationary points for the field equations and investigating their stability
properties. This approach has been widely studied in various theories and it
has provided important information for cosmological evolution
\cite{dn1,dn2,dn3,dn4,dn5,dn6}. Homogeneous and anisotropic spacetimes have
not been widely studied in the literature. Such an analysis is important in
order to understand the dynamical effects of scalar-torsion in the very early
universe. The structure of the paper is as follows.

In Section \ref{sec2} we present the basic properties and definition of
teleparallelism, and we give the Action Integral of the gravitational theory
that we discuss, the scalar-torsion theory. The Kantowski-Sachs geometry is
considered in Section \ref{sec3}. We select a specific frame for the vierbein
fields in which the limit of General Relativity is recovered. The field
equations are determined. Moreover, we rewrite the field equations with the
use of a new set of dimensional variables. Hence, the field equations are
written in the equivalent form of an algebraic-differential system. The
dynamical analysis is presented for two forms of the scalar field potential,
the exponential potential and the power-law potential. The results are given
in Sections \ref{sec4} and \ref{sec5}, respectively. Finally, in \ref{sec6} we
summarize our results and we draw our conclusions.

\section{Teleparallel theory of gravity}

\label{sec2}

Consider now the vierbein fields ${\mathbf{e}_{\mu}(x^{\sigma})}$ \ with
commutator relations $[e_{\mu},e_{\nu}]=c_{\nu\mu}^{\beta}e_{\beta}~$\ , where
$c_{\left(  \nu\mu\right)  }^{\beta}=0.$ The\ vierbein fields form a basis for
the tangent space at each point $P$ with the relation $(e_{\mu},e_{\nu
})=\mathbf{e}_{\mu}\cdot\mathbf{e}_{\nu}=g_{\mu\nu}$. In the nonholonomic
coordinates we consider the covariant derivative $\nabla_{\mu}$ defined with
the connection%
\begin{equation}
\mathring{\Gamma}_{\nu\beta}^{\mu}=\{_{\nu\beta}^{\mu}\}+\mathring{\Gamma
}_{\nu\beta}^{\mu},
\end{equation}
where $\{_{\nu\beta}^{\mu}\}$ is the symmetric Levi-Civita and
\begin{equation}
\mathring{\Gamma}_{\nu\beta}^{\mu}=\frac{1}{2}g^{\mu\sigma}(c_{\nu\sigma
,\beta}+c_{\sigma\beta,\nu}-c_{\mu\beta,\sigma}), \label{sd1}%
\end{equation}
where $\mathring{\Gamma}_{\nu\beta}^{\mu}$ are antisymmetric in the two first
indices, that is $~\mathring{\Gamma}_{\mu\nu\beta}=-\mathring{\Gamma}_{\nu
\mu\beta},~\mathring{\Gamma}_{\mu\nu\beta}=\eta_{\mu\sigma}\mathring{\Gamma
}_{\nu\beta}^{\mu},$ and it is related to the commutator relations of the
vierbein fields.

In teleparallelism the Weitzenb{\"{o}}ck connection is considered
\cite{Weitzenb23}, which means that the connection defines the flat space,
i.e., $\mathbf{e}_{\mu}\cdot\mathbf{e}_{\nu}=\eta_{\mu\nu}$, where now
definition (\ref{sd1}) is
\begin{equation}
\mathring{\Gamma}_{\nu\beta}^{\mu}=\frac{1}{2}\eta^{\mu\sigma}(c_{\nu
\sigma,\beta}+c_{\sigma\beta,\nu}-c_{\mu\beta,\sigma}).
\end{equation}

We define the nonnull torsion tensor from the relation
\begin{equation}
T_{\mu\nu}^{\beta}=2\mathring{\Gamma}_{\left[  \nu\mu\right]  }^{\beta}%
\end{equation}
and the torsion scalar \cite{Hayashi79}%
\begin{equation}
T=\frac{1}{2}({K^{\mu\nu}}_{\beta}+\delta_{\beta}^{\mu}{T^{\theta\nu}}%
_{\theta}-\delta_{\beta}^{\nu}{T^{\theta\mu}}_{\theta}){T^{\beta}}_{\mu\nu}%
\end{equation}
in which $K_{~~~\beta}^{\mu\nu}=-\frac{1}{2}({T^{\mu\nu}}_{\beta}-{T^{\nu\mu}%
}_{\beta}-{T_{\beta}}^{\mu\nu})$ is the contorsion tensor that equals the
difference between the connections in the holonomic and the non-holonomic frame.

The fundamental Action Integral in teleparallelism is \cite{cc}
\begin{equation}
S_{T}=\frac{1}{16\pi G}\int d^{4}xeT~+S_{m},~e=\det(e_{\mu}), \label{cc.05}%
\end{equation}
where $e=\det(e_{\mu})$ is the determinant of the vierbein fields and $S_{m}$
is the Action Integral corresponding to the matter source. The matter source
can be a dust fluid, radiation, Chaplygin gas, the cosmological constant,
scalar field and many others. In the latter cases, the gravitational field
equations which follow from the variation of (\ref{cc.05}) are equivalent with
those of General Relativity.

\subsection{Scalar-torsion theory}

Inspired by the scalar-tensor theory which is defined by Mach's principle,
consider now a nonminimally coupled scalar field $\phi$ which interact with
the Lagrangian of TEGR. The resulting Machian teleparallel gravitational
theory is defined by the Action Integral \cite{ss3}%
\begin{equation}
S=\frac{1}{16\pi G}\int d^{4}xe\left[  \hat{F}\left(  \psi\right)  T+\frac
{1}{2}\psi_{;\mu}\psi^{\mu}+\hat{V}\left(  \psi\right)  \right]  \label{d.08a}%
\end{equation}
and it is known as scalar-torsion theory. $\hat{V}\left(  \psi\right)  $ is
the scalar field potential and $\hat{F}\left(  \psi\right)  $ is the coupling
function of the scalar field with the torsion scalar.

Without loss of generality we can perform a change of variables $\psi
\rightarrow\phi$, such that we can write the Action Integral (\ref{d.08a}) in
the following form
\begin{equation}
S=\frac{1}{16\pi G}\int d^{4}xe\left[  F\left(  \phi\right)  \left(
T+\frac{\omega}{2}\phi_{;\mu}\phi^{\mu}+V\left(  \phi\right)  \right)
\right]  , \label{d.08}%
\end{equation}
$\hat{V}\left(  \psi\right)  =F\left(  \phi\right)  V\left(  \phi\right)  $,
and $d\psi=\sqrt{\omega F\left(  \phi\right)  }d\phi$. The parameter $\omega$
is a constant parameter similar to the Brans-Dicke parameter \cite{ks8}. From
(\ref{d.08}) we conclude that scalar-torsion theory is a second-order theory
of gravity.

Scalar-tensor theory is equivalent with a scalar field in Einstein-frame under
the application of a conformal transformation. That property is not true for
the scalar-torsion theory. Consequently, the scalar-tensor and scalar-torsion
theories are not related through a conformal transformation. It is known that
the additional degrees of freedom in $f\left(  R\right)  $-gravity can be
attributed to a scalar field non-minimally coupled to the Ricciscalar, the
equivalent higher-order theory in teleparallelism is the $f\left(  T,B\right)
$-gravity, where $B$ is the boundary term and differentiates the torsion $T$
and the Ricci scalar \cite{ss2a,ss3a}.

\section{Kantowski-Sachs cosmology}

\label{sec3}

The Kantowski-Sachs geometry in the Misner-like variables is described by the
line element
\begin{equation}
ds^{2}=-N^{2}\left(  t\right)  dt^{2}+e^{2\alpha\left(  t\right)  }\left(
e^{2\beta\left(  t\right)  }dx^{2}+e^{-\beta\left(  t\right)  }\left(
dy^{2}+f^{2}\left(  y\right)  dz^{2}\right)  \right)  , \label{ch.03}%
\end{equation}
where $N\left(  t\right)  $ is the lapse function, $\alpha\left(  t\right)  $
is the scale factor which describes the size of the three-dimensional
hypersurface and $\beta\left(  t\right)  $ is the anisotropic parameter.

The definition of the proper vierbein fields is essential in order that the
limit of General Relativity be recovered. If we follow the definition given in
\cite{err1,err2}, we see that the limit of General Relativity does not exist.

We follow \cite{ss10} and we assume the vierbein fields%
\begin{align*}
e^{1}  &  =Ndt~,\\
e^{2}  &  =e^{a+\beta}\cos z\sin y~dx+e^{a-\frac{\beta}{2}}\left(  \cos y\cos
z~dy-\sin y\sin z~dz\right)  ~,\\
e^{3}  &  =e^{a+\beta}\sin y\sin z~dx+e^{a-\frac{\beta}{2}}\left(  \cos y\sin
z~dy-\sin y\cos z~dz\right)  ~,\\
e^{4}  &  =e^{a+\beta}\cos y~dx-e^{a-\frac{\beta}{2}}\sin y~dy.
\end{align*}

Hence, the torsion scalar for the line element (\ref{ch.03}) with the use of
these vierbein fields is calculated%
\begin{equation}
T=\frac{1}{N^{2}}\left(  6\dot{\alpha}^{2}-\frac{3}{2}\dot{\beta}^{2}\right)
-2e^{-2\alpha+\beta}. \label{ch.03A}%
\end{equation}

Thus, with the use of (\ref{ch.03A}) and by assuming that the scalar field
inherits the symmetries of the background space the Action Integral
(\ref{d.08}) is%
\begin{equation}
S=\frac{1}{16\pi G}\int dt\left(  F\left(  \phi\right)  e^{3\alpha}\left(
\frac{1}{N}\left(  6\dot{\alpha}^{2}-\frac{3}{2}\dot{\beta}^{2}-\frac{\omega
}{2}\dot{\phi}^{2}\right)  +N\left(  V\left(  \phi\right)  -2e^{-2\alpha
+\beta}\right)  \right)  \right)  . \label{ch.03B}%
\end{equation}

By variation of the Action Integral (\ref{ch.03B}) with respect to the
dynamical variables $N,~\alpha,~\beta$ and $\phi$ we derive the field
equations. They are
\begin{equation}
0=\ddot{\alpha}+\frac{3}{2}\dot{\alpha}^{2}+\frac{3}{8}\dot{\beta}^{2}%
+\frac{1}{4}\left(  \frac{\omega}{2}\dot{\phi}^{2}-V\left(  \phi\right)
\right)  +\frac{d}{dt}\left(  \ln F\left(  \phi\right)  \right)  \dot{\alpha
}+\frac{1}{6}e^{-2\alpha+\beta}~, \label{mm.01}%
\end{equation}%
\begin{equation}
0=\ddot{\beta}+3\dot{\alpha}\dot{\beta}+\frac{d}{dt}\left(  \ln F\left(
\phi\right)  \right)  \dot{\beta}-\frac{2}{3}e^{-2\alpha+\beta}~,
\label{mm.02}%
\end{equation}%
\begin{equation}
0=\omega\left(  \ddot{\phi}+3\dot{\alpha}\dot{\phi}\right)  \dot{\phi}+\dot
{V}+\frac{d}{dt}\left(  \ln F\left(  \phi\right)  \right)  \left(  6\dot
{a}^{2}-\frac{3}{2}\dot{\beta}^{2}+V\left(  \phi\right)  -2e^{-2\alpha+\beta
}\right)  ~ \label{mm.03}%
\end{equation}
and the constraint
\begin{equation}
0=F\left(  \phi\right)  e^{3\alpha}\left(  6\dot{\alpha}^{2}-\frac{3}{2}%
\dot{\beta}^{2}-\frac{\omega}{2}\dot{\phi}^{2}-V\left(  \phi\right)
+2e^{-2\alpha+\beta}\right)  , \label{mm.04}%
\end{equation}
where without loss of generality we have selected $N\left(  t\right)  =1$.
Recall that in the Misner variables the Hubble function is $H=\dot{a}$ and the
shear $\sigma^{2}\simeq\dot{\beta}^{2}$

We define the following set of new variables
\[
\Sigma=\frac{\dot{\beta}}{2D}~,~x=\frac{\dot{\phi}}{12D}~,~\ y=\sqrt
{\frac{V\left(  \phi\right)  }{6D^{2}}}~,
\]%
\[
\eta=\frac{H}{D}~,~D=\sqrt{H^{2}+\frac{1}{3}e^{-2\alpha+\beta}~,}%
\]
where $D$ is the normalization parameter. Hence, the Hubble function is
$H=e^{\frac{\beta}{2}-\alpha}\frac{\eta}{\sqrt{1-\eta^{2}}}$, from which we
infer that $\eta^{2}\leq1$. In the limit $\eta^{2}=1$, it follows that
$H^{2}+\frac{1}{3}e^{-2\alpha+\beta}\simeq H^{2}$, that is, the asymptotic
solution has a spatially flat three-dimensional hypersurface. Hence, the
Kantowski-Sachs geometry is reduced to a Bianchi I spacetime.

In the new variables $\left\{  \Sigma,x,y,\eta,\lambda\right\}  $ and for the
new dependent variable $d\tau=Ddt$, the field equations (\ref{mm.01}%
)-(\ref{mm.03}) are expressed as follow%
\begin{equation}
\frac{d\Sigma}{d\tau}=1-\Sigma\left(  4\sqrt{3}x+\Sigma\right)  -\frac{3}%
{2}\eta\Sigma\left(  1-\omega x^{2}+y^{2}-\Sigma^{2}\right)  +\eta^{2}\left(
4\sqrt{3}x\Sigma+\Sigma^{2}-1\right)  , \label{mm.05}%
\end{equation}%
\begin{align}
\frac{dx}{d\tau}  &  =x^{2}\left(  \frac{\omega}{2}x\eta+4\sqrt{3}\left(
2\eta^{2}-1\right)  \right)  -\frac{2\sqrt{3}}{\omega}\left(  \left(
2+\lambda\right)  y^{2}+2\left(  2\eta^{2}-\Sigma^{2}\right)  \right)
\label{mm.06}\\
&  +x\left(  3\eta\Sigma^{2}-3\eta\left(  1+y^{2}\right)  -2\left(  1-\eta
^{2}\right)  \right)  ,\nonumber
\end{align}%
\begin{equation}
\frac{dy}{d\tau}=\frac{1}{2}y\left(  x\left(  3\omega x\eta+2\sqrt{3}\left(
\lambda+4\eta^{2}\right)  -2\Sigma+\eta\left(  3\left(  1-y^{2}\right)
+2\eta\Sigma+3\Sigma^{2}\right)  \right)  \right)  , \label{mm.07}%
\end{equation}%
\begin{equation}
\frac{d\eta}{d\tau}=\frac{1}{2}\left(  \eta^{2}-1\right)  \left(  1+3\left(
\omega x^{2}-y^{2}+\Sigma^{2}\right)  +2\eta\left(  4\sqrt{3}x+\Sigma\right)
\right)  \label{mm.08}%
\end{equation}
with the algebraic constraint
\begin{equation}
1-\omega x^{2}-y^{2}-\Sigma^{2}=0. \label{mm.09}%
\end{equation}

The parameter $\lambda$ is defined as $\lambda=\left(  \ln V\left(
\phi\right)  \right)  _{,\phi}$. For the exponential potential, $V\left(
\phi\right)  =V_{0}e^{\lambda_{0}\phi}$, it follows that $\lambda=\lambda_{0}%
$, which means that $\lambda$ is always a constant parameter. However for
other potential functions $\lambda$ varies in terms of time. In particular, we
find that%
\begin{equation}
\frac{d\lambda}{d\tau}=2\sqrt{3}x\lambda^{2}\left(  \Gamma\left(
\lambda\right)  -1\right)  \text{,~}\Gamma\left(  \lambda\left(  \phi\right)
\right)  =\frac{V_{,\phi\phi}V}{\left(  V_{,\phi}\right)  ^{2}}. \label{mm.10}%
\end{equation}

The cosmological parameters in the new variables are expressed as follow%
\begin{equation}
w_{eff}\left(  \Sigma,x,y,\eta\right)  =\frac{1}{3\eta^{2}}\left(  1+3\left(
\omega x^{2}-y^{2}+\Sigma^{2}\right)  +\eta\left(  8\sqrt{3}x-\eta\right)
\right)  ,
\end{equation}
and%
\begin{equation}
q\left(  \Sigma,x,y,\eta\right)  =\frac{4\eta^{2}-3}{2\eta^{4}}\left(
1+3\left(  \omega x^{2}-y^{2}+\Sigma^{2}\right)  +\eta\left(  8\sqrt{3}%
x-\eta\right)  \right)
\end{equation}
in which $w_{eff}\left(  \Sigma,x,y,\eta\right)  $ is the equation of state
parameter for the effective fluid and $q\left(  \Sigma,x,y,\eta\right)  $ is
the deceleration parameter.

\section{Dynamical analysis for the exponential potential}

\label{sec4}

In the following we study the asymptotic behaviour of the dynamical parameters
described by the field equations (\ref{mm.05})-(\ref{mm.08}) in which for the
scalar field potential we consider the exponential function, $V\left(
\phi\right)  =V_{0}e^{\lambda\phi}$.

We determine all the stationary points of the phase-space and we study their
stability properties. From the latter we reconstruct the cosmological history
and we get important information for the initial value problem. Each
stationary point describes an asymptotic solution for the field equations
which corresponds to a specific epoch of the cosmological evolution.

Let $\frac{d}{d\tau}\mathbf{X}=\mathbf{F}\left(  \mathbf{X}\right)  $ be the
dynamical system (\ref{mm.05})-(\ref{mm.08}), $\mathbf{X}=\left(
\Sigma,x,y,\eta\right)  ^{T}$. Consider the point $P$ which satisfies the
condition $\mathbf{F}\left(  \mathbf{X}\left(  P\right)  \right)  =0$. Then
$P$ is a stationary/critical point for the dynamical system. At point $P$, we
can calculate the physical parameters $w_{eff}\left(  P\right)  $ and
$q\left(  P\right)  $ in order to infer the physical properties of the
asymptotic solution. For instance, when $q\left(  P\right)  <0$, the
asymptotic solution describes an accelerated universe, while, if
$\Sigma\left(  P\right)  =0$, the asymptotic solution describes an isotropic solution.

In order to determine the stability properties of the dynamical system near to
the stationary point $P$, we perform a linearization of the dynamical system
near to the point $P$ and we calculate the eigenvalues of the linearized
matrix. The point $P$ is an attractor when all the eigenvalues have negative
real parts. Point $P$ is characterized as a source when all the eigenvalues
have positive real parts, otherwise is a saddle point. The knowledge of the
stability properties is necessary in order to infer the physical properties of
the gravitational theory that we study, as also to get constraints for the
initial conditions.

The dynamical variables of the field equations (\ref{mm.05})-(\ref{mm.08})
satisfy the constraint equation (\ref{mm.09}). Hence, with the use of the
constraint the dimension of the phase-space is reduced by one. We remark that
parameter $y$ is always positive. For $\omega>0$ the constraint equation
(\ref{mm.09}), the dynamical variables take values on a three-dimensional
ellipsoid, which reduces to a sphere for $\omega=1$, which means that
variables $\left(  \Sigma,x,y\right)  $ are defined in the finite regime. On
the other hand for $\omega<0$ the dynamical variables are not constrainted
which means that we should investigate the existence of stationary points at
the infinity regime.

\subsection{Stationary points at the finite regime}

At the finite regime the stationary points $P$ of the dynamical system which
satisfy the algebraic constraint (\ref{mm.09}) are%
\begin{equation}
P_{1}^{\pm}=\left(  \pm\sqrt{1-\omega\left(  x_{1}\right)  ^{2}}%
,x_{1},0,1\right)  ,
\end{equation}
where $x_{1}$ is arbitrary and $1-\omega\left(  x_{1}\right)  ^{2}\geq0$ in
order that the points be real and physically accepted. For $\omega>0$,
$\left(  x_{1}\right)  ^{2}\leq\frac{1}{\omega}$, while for $\omega<0$ points
$P_{1}^{\pm}$ are always real. Points $P_{1}^{\pm}$ describe Kasner-like
anisotropic solutions of Bianchi I geometry with $q\left(  P_{1}^{\pm}\right)
=2+4\sqrt{3}x_{1}$.
\begin{equation}
P_{2}^{\pm}=\left(  \pm\sqrt{1-\omega\left(  x_{2}\right)  ^{2}}%
,x_{2},0,-1\right)  ,
\end{equation}
in which $x_{2}$ is arbitrary and constrained by the algebraic relation
$1-\omega\left(  x_{2}\right)  ^{2}\geq0$. Similarly to above for $\omega>0$,
$\left(  x_{2}\right)  ^{2}\leq\frac{1}{\omega}$, while for $\omega<0$
points$~P_{2}^{\pm}$ are always real. The asymptotic solutions are those of
Kasner-like anisotropic spacetimes with $q\left(  P_{2}^{\pm}\right)
=2-4\sqrt{3}x_{2}.$%
\[
P_{3}^{\pm}=\left(  \mp\sqrt{\frac{\omega}{\omega+48}},\mp\frac{4\sqrt{3}%
}{\sqrt{\omega\left(  \omega+48\right)  }},0,\pm2\sqrt{\frac{\omega}%
{\omega+48}}\right)
\]
$P_{3}^{\pm}~$are real points and the asymptotic solutions are physically
accepted for $\omega\left(  \omega+48\right)  >0$; that is, $\omega>0$ or
$\omega<-48.$However, from the definition of $\eta^{2}\leq1$ it follows
$0<\omega\leq16.$ Hence, the points exist only for positive $\omega$. The
asymptotic solutions at the points $P_{3}^{\pm}$ describe anisotropic
Kantowski-Sachs geometries with $w_{eff}\left(  P_{3}^{\pm}\right)  =0$ and
$q\left(  P_{3}^{\pm}\right)  =\frac{13}{8}-\frac{18}{\omega}$.\ We conclude
that the asymptotic solutions at points $P_{3}^{\pm}~$describe accelerated
universes for $0<\omega<\frac{144}{13}$.

\subsubsection{Stability analysis}

We now proceed with the stability analysis for the stationary points. We make
use of the algebraic constraint (\ref{mm.09}) and we reduce by one the
dimension of the phase-space, such that the stationary points are on the three
dimensional manifold of the variables $\left\{  \Sigma,x,\eta\right\}  $.

The eigenvalues of the linearized dynamical system near to the stationary
points $P_{1}^{\pm}$ are
\[
e_{1}\left(  P_{1}^{\pm}\right)  =0~,~e_{2}\left(  P_{1}^{\pm}\right)
=6+2\sqrt{3}\left(  4+\lambda\right)  x_{1}~,~e_{3}\left(  P_{1}^{\pm}\right)
=2\left(  2+4\sqrt{3}x_{1}\pm\sqrt{1-\omega\left(  x_{1}\right)  ^{2}}\right)
.
\]

Similarly, for the points $P_{2}^{\pm}$ the eigenvalues are derived to be
\[
e_{1}\left(  P_{2}^{\pm}\right)  =0~,~e_{2}\left(  P_{2}^{\pm}\right)
=-6+2\sqrt{3}\left(  4+\lambda\right)  x_{2}~,~e_{3}\left(  P_{2}^{\pm
}\right)  =2\left(  -2+4\sqrt{3}x_{1}\pm\sqrt{1-\omega\left(  x_{2}\right)
^{2}}\right)  .
\]

These two sets of points have at least one eigenvalue with zero real part,
that is eigenvalue $e_{1}$, while there are ranges of the free parameters for
which the eigenvalues $e_{2}$ and $e_{3}$ can have negative real parts. The
latter means that there may exist a stable submanifold in the dynamical
system. However, the derivation of this submanifold is of mathematical
interest and does not contribute in the physical discussion. Thus, the
stability properties for these two set of points are investigated numerically.

\begin{figure}[ptb]
\centering\includegraphics[width=1\textwidth]{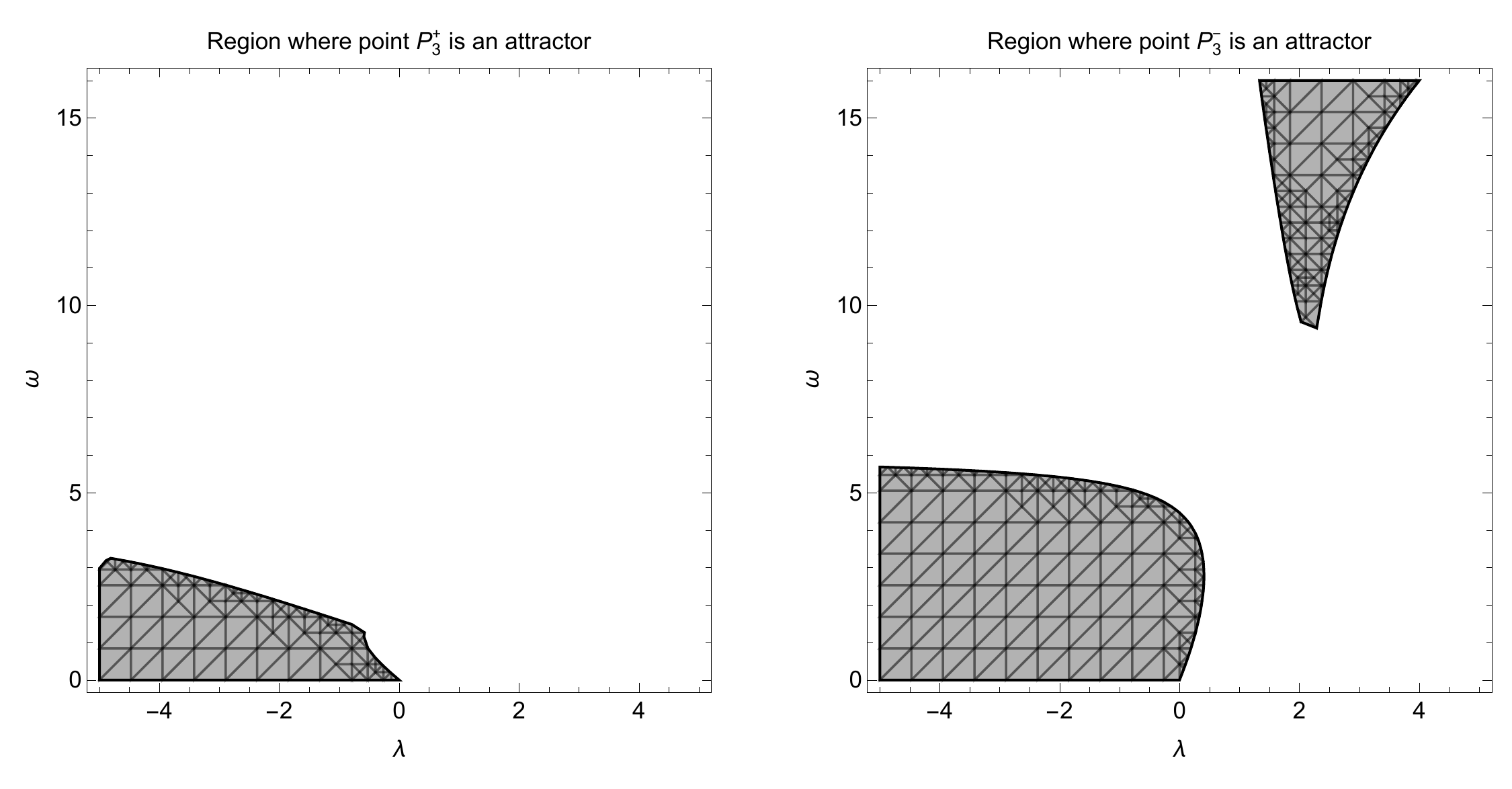}\caption{Region plot in
the space $\left(  \lambda,\omega\right)  $ the stationary points $P_{3}^{\pm
}$ are attractors and the Kantowski-Sachs solutions are stable. }%
\label{fig01}%
\end{figure}

For the points $P_{3}^{\pm}$ the eigenvalues are determined numerically. In
Fig \ref{fig01} we present the region plots in the space of the free variables
$\left(  \lambda,\omega\right)  $ for which the points have all the
eigenvalues with negative real parts.

\begin{figure}[ptb]
\centering\includegraphics[width=1\textwidth]{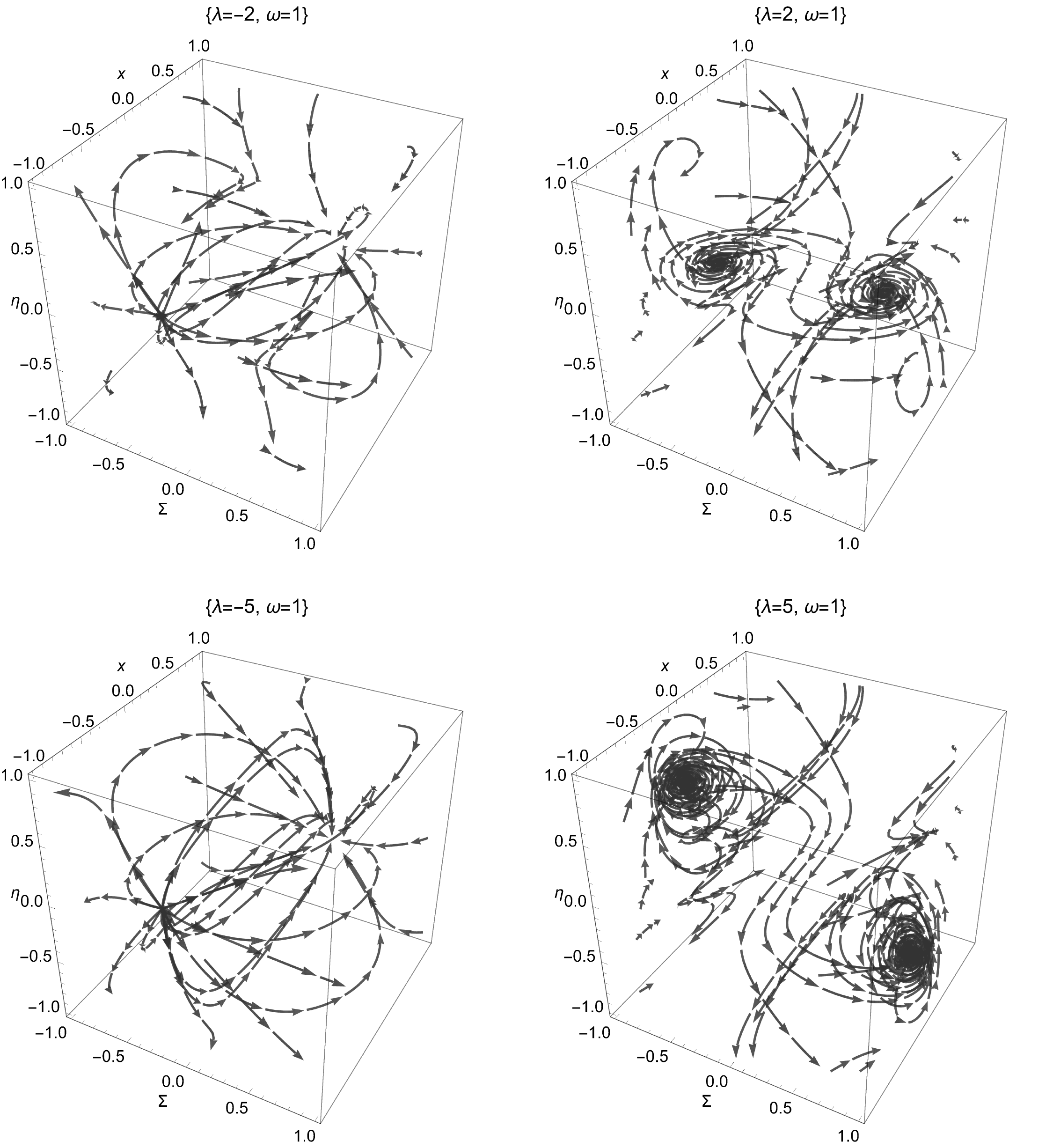}\caption{Phase-space
portraits for the three-dimensional dynamical system in the space $\left(
\Sigma,x,\eta\right)  $ for the Kantowski-Sachs spacetime and for $\omega=1$.
In the first row the left fig. is for $\lambda=-2$ and the right fig. is for
$\lambda=2$. In the second row the left fig. is for $\lambda=-5$ and the right
fig. is for $\lambda=-2$. We observe that points $P_{3}^{\pm}$ can be the
attractors in the finite regime. }%
\label{fig02}%
\end{figure}

\begin{figure}[ptb]
\centering\includegraphics[width=1\textwidth]{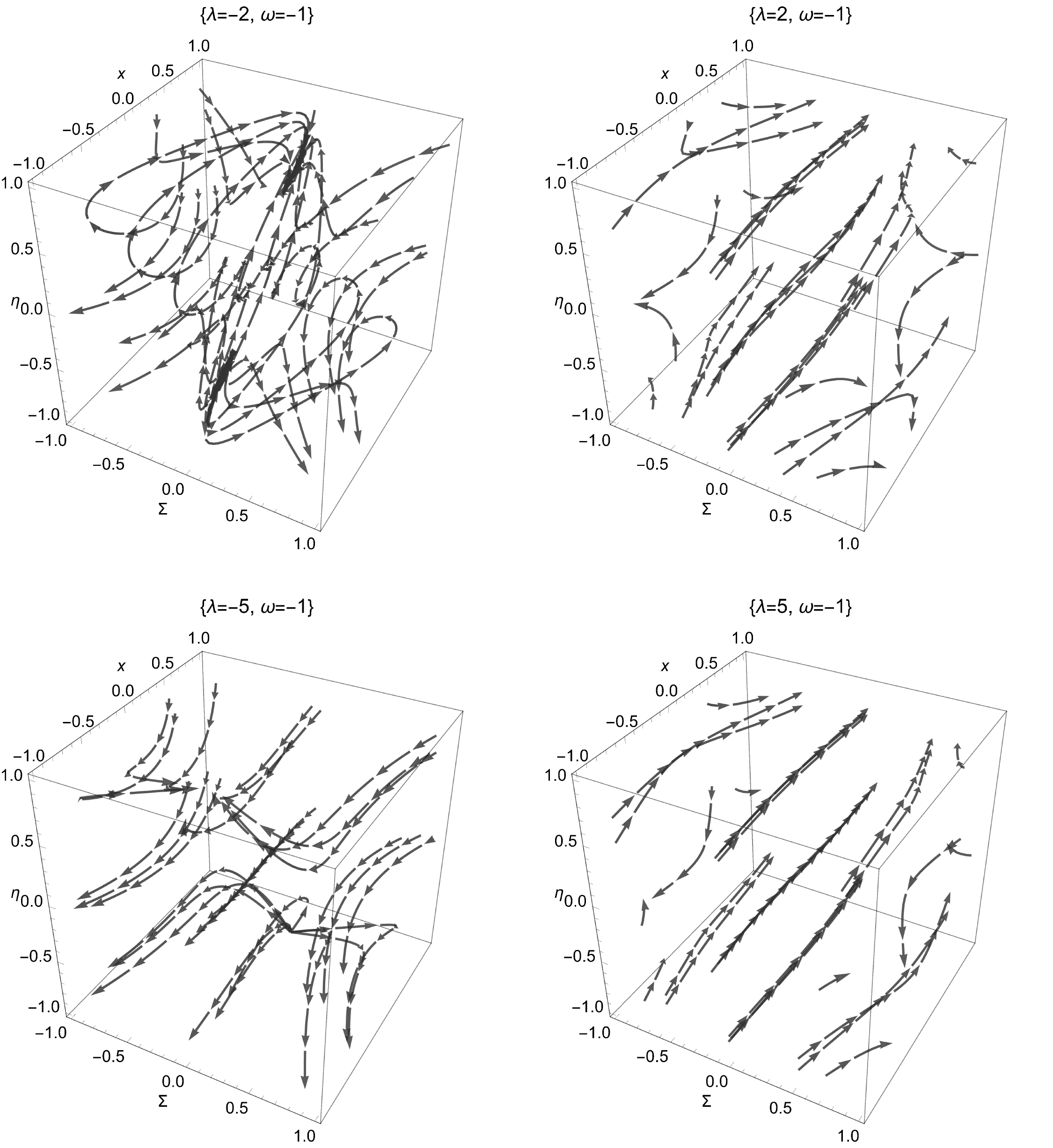}\caption{Phase-space
portraits for the three-dimensional dynamical system in the space $\left(
\Sigma,x,\eta\right)  $ for the Kantowski-Sachs spacetime and for $\omega=-1$.
In the first row the left fig. is for $\lambda=-2$ and the right fig. is for
$\lambda=2$. In the second row the left fig. is for $\lambda=-5$ and the right
fig. is for $\lambda=-2$. We observe that the trajectories move to infinity.}%
\label{fig03}%
\end{figure}

In the series of Figs. \ref{fig02} and \ref{fig03} we present the evolution of
the trajectories for the field equations in the three dimensional space
$\left(  \Sigma,x,\eta\right)  $. In Fig. \ref{fig02} the plots are for
positive value $\omega$, where we observe that points $P_{3}^{\pm}$ can be
attractors. On the other hand Fig. \ref{fig03} is for $\omega<0$ there are no
attractors in the finite regime and the trajectories move to infinity.
Finally, for the points $P_{1}^{\pm}$ and $P_{2}^{\pm}$ from the numerical
simulations we can conclude that the stationary points do not describe stable solutions.

\subsection{Stationary points at the infinity}

Consider now the case for which $\omega<0$. In order to investigate the
existence of stationary points at the infinity regime, we consider the new set
of Poincare variables
\[
\Sigma=\frac{\rho}{\sqrt{1-\rho^{2}}}\sin\theta~,~x=\frac{\rho}{\sqrt
{1-\rho^{2}}}\cos\theta~,~\omega=-\Omega^{2}\text{}%
\]
with $\rho\in\left[  0,1\right]  $ and $\theta\in\lbrack0,2\pi)$.

Moreover, we define the new independent variable, $dT=\sqrt{1-\rho^{2}}d\tau$,
such that the field equations are%
\begin{align}
4\Omega^{2}\frac{d\rho}{dT}  &  =-6\Omega^{2}\eta\rho\sqrt{1-\rho^{2}}\left(
2+\left(  \Omega^{2}-3+\left(  1+\Omega^{2}\right)  \cos\left(  2\theta
\right)  \right)  \rho^{2}\right) \nonumber\\
&  +\left(  1-\rho^{2}\right)  \left(  4\left(  \sqrt{3\lambda}\cos
\theta+\Omega^{2}\sin\theta\right)  +\left(  \sqrt{3}\left(  \Omega^{2}\left(
3\lambda-4\right)  -\left(  4+5\lambda\right)  \right)  \cos\theta\right)
\rho^{2}\right)  +\nonumber\\
&  +\left(  1-\rho^{2}\right)  \rho^{2}\left(  \sqrt{3}\left(  4+\lambda
\right)  \left(  1+\Omega^{2}\right)  \cos\left(  3\theta\right)  -8\Omega
^{2}\sin\theta\right)  +\nonumber\\
&  +4\eta^{2}\left(  1-\rho^{2}\right)  \left(  \Omega^{2}\sin\theta\left(
2\rho^{2}-1\right)  +4\sqrt{3}\cos\theta\left(  1+\left(  \Omega^{2}-1\right)
\rho^{2}\right)  \right)  ,
\end{align}%
\begin{align}
\frac{\Omega^{2}\rho}{\sin\theta}\frac{d\theta}{dT}  &  =\left(  1-\rho
^{2}\right)  \left(  \Omega^{2}\cot\theta-\left(  4\sqrt{3}+\Omega^{2}%
\cot\theta\right)  \eta^{2}\right) \nonumber\\
&  +\sqrt{3}\left(  \left(  \lambda-\left(  4+\lambda\right)  \left(
\Omega^{2}\cos^{2}\theta-\sin^{2}\theta\right)  \right)  \rho^{2}%
-\lambda\right)  ,
\end{align}%
\begin{align}
\frac{d\eta}{dT}  &  =\left(  1-\eta^{2}\right)  \left(  -\left(  4\sqrt
{3}\cos\theta+\sin\theta\right)  \eta\rho\right) \nonumber\\
&  +\frac{\left(  1-\eta^{2}\right)  }{\sqrt{1-\rho^{2}}}\left(  1+\left(
3\left(  \Omega^{2}\cos^{2}\theta-\sin^{2}\theta\right)  -1\right)  \rho
^{2}\right)  . \label{in.03}%
\end{align}

Infinity is reached when $\rho\rightarrow1$. However, from (\ref{in.03}) we
observe that stationary points exist only on the surfaces with $\eta^{2}=1$,
or $\eta=0$ and $\left(  1+\left(  3\left(  \Omega^{2}\cos^{2}\theta-\sin
^{2}\theta\right)  -1\right)  \rho^{2}\right)  =0$.

For $\eta^{2}=1$ and $\rho\rightarrow1$, the field equations become%
\begin{equation}
\frac{d\rho}{dT}=0~,~\frac{d\eta}{dT}=0
\end{equation}%
\begin{equation}
\frac{d\theta}{dT}=-\frac{\sqrt{3}}{2\Omega^{2}}\sin\theta\left(  \Omega
^{2}-1+\left(  1+\Omega^{2}\right)  \cos2\theta\right)  .
\end{equation}

Consequently, stationary points exist for $\theta_{1}=0~$\ or$~\theta
_{2}=\frac{1}{2}\arccos\frac{1-\Omega^{2}}{1+\Omega^{2}}$. The point with
$\theta_{1}=0$ describes an isotropic spatially flat\ FLRW spacetime, while
the asymptotic solution at $\theta_{2}=\frac{1}{2}\arccos\frac{1-\Omega^{2}%
}{1+\Omega^{2}}$ is a Kasner-like solution in Bianchi I geometry.

One of the eigenvalues of the linearized system at the isotropic solutions has
always positive real part, while one eigenvalue has always negative real part,
consequently, the stationary point is a saddle point and the solution is
unstable. On the other hand, for the Bianchi I point with $\theta_{2}=\frac
{1}{2}\arccos\frac{1-\Omega^{2}}{1+\Omega^{2}}$ one of the eigenvalues is
found to be always positive, which means that the point describes always an
unstable solution.

The surface with $\eta=0$ describe static spacetimes. The stationary points
are $\theta_{3}=0~$and$~\theta_{4}=\frac{1}{2}\arccos\frac{1-\Omega^{2}%
}{1+\Omega^{2}}$ with similar physical properties as before. Indeed, for
$\theta=\theta_{3}$ the asymptotic solution is the static isotropic closed
FLRW spacetime, while for $\theta=\theta_{4}$ the asymptotic solution
describes the static Kantowski-Sachs universe. The stability properties of
these stationary points are investigated numerically, from where it follows
that the static asymptotic solutions are always unstable.

\begin{figure}[ptb]
\centering\includegraphics[width=1\textwidth]{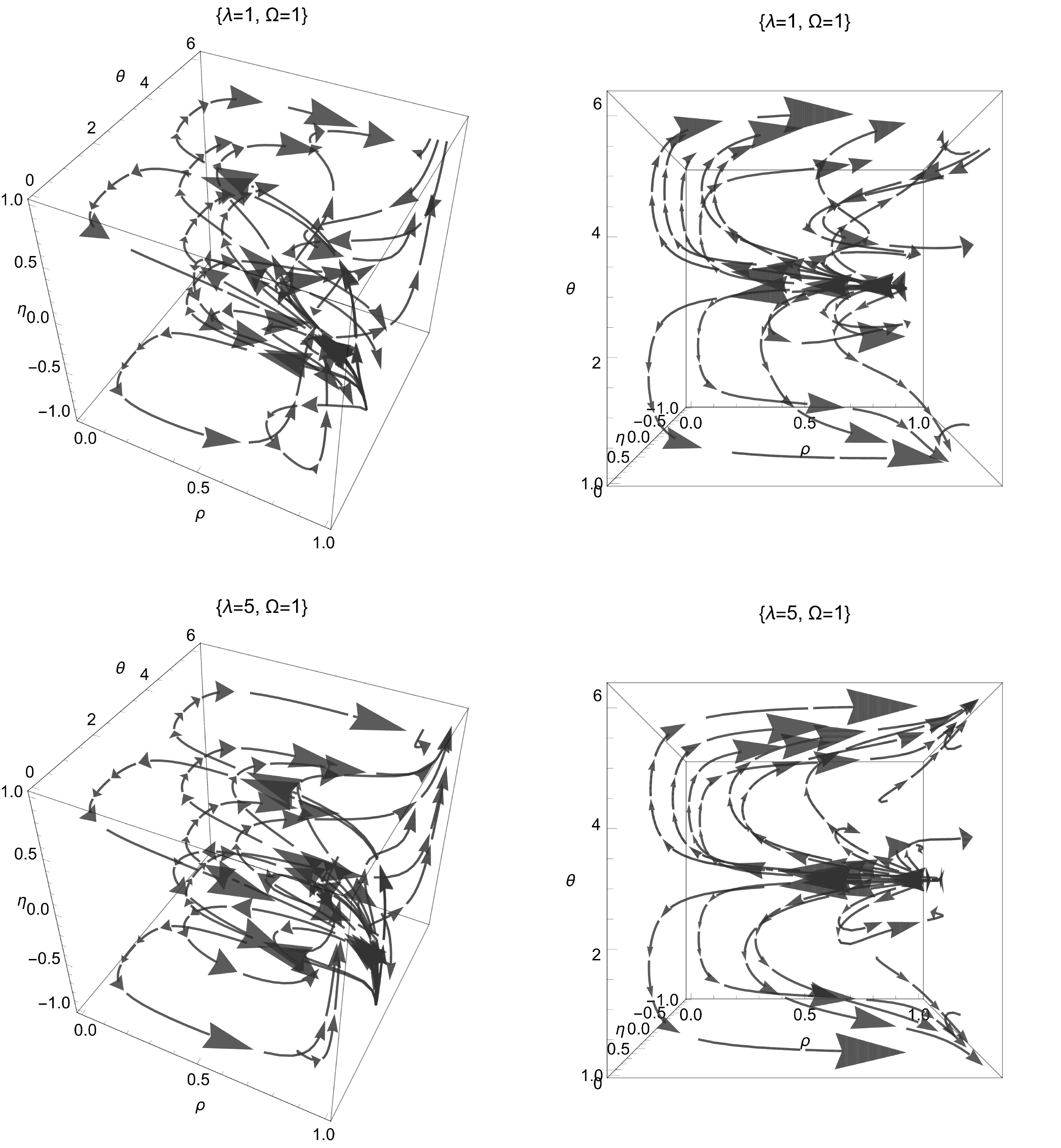}\caption{Phase-space
portraits for the three-dimensional dynamical system in the space $\left(
\rho,\theta,\eta\right)  $ for the Kantowski-Sachs spacetime and for
$\omega=-1,$ that is $\Omega=1$. First row are plots for $\lambda=1$ while in
the second row are plots with $\lambda=5.$ The plots of the left and right
columns are the same but from different view point. It is clear that in there
are not attractos for the dynamical system and the trajectories start from the
finite regime, reach infinity and vice-versa. }%
\label{fig04}%
\end{figure}

\begin{figure}[ptb]
\centering\includegraphics[width=1\textwidth]{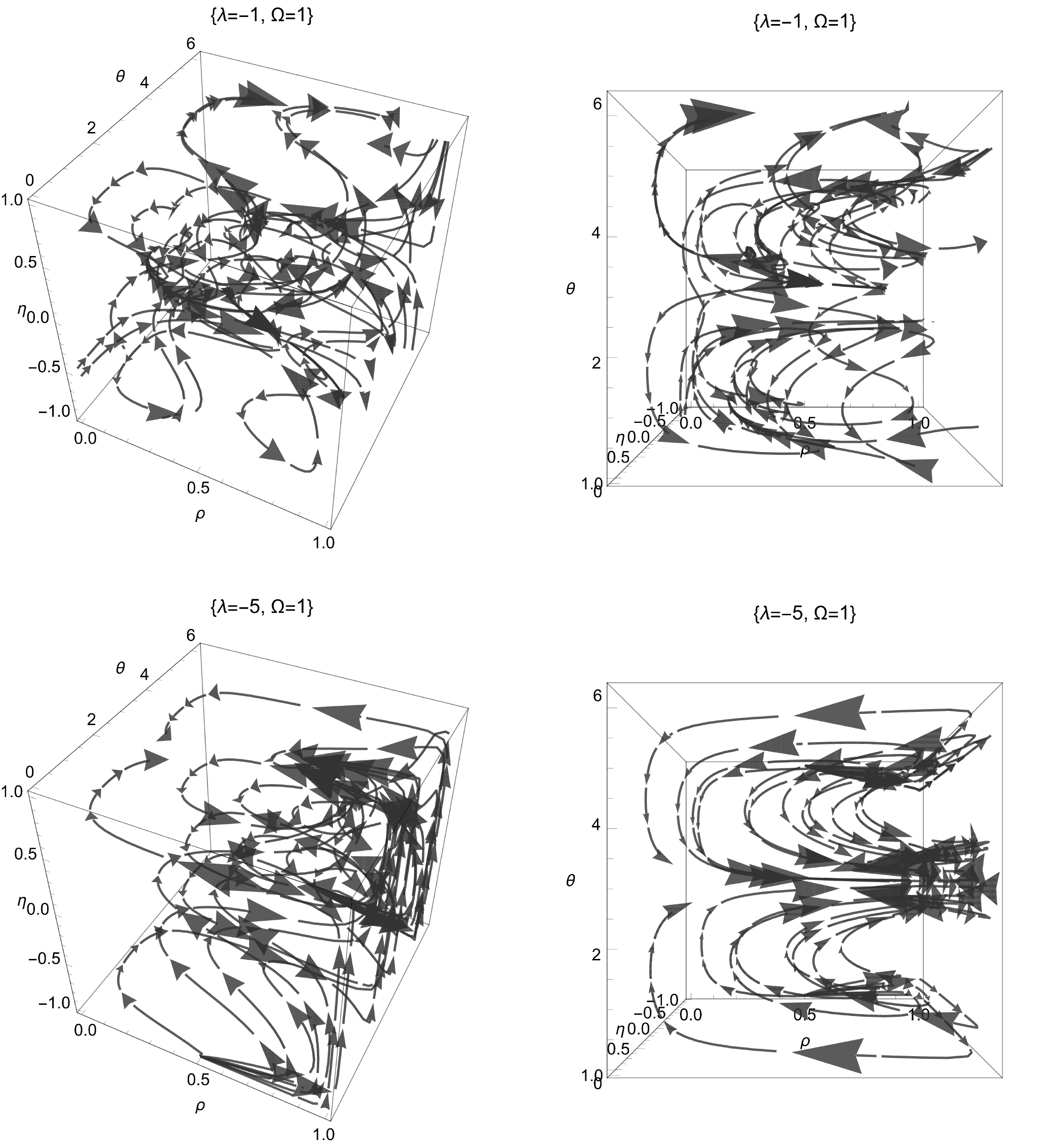}\caption{Phase-space
portraits for the three-dimensional dynamical system in the space $\left(
\rho,\theta,\eta\right)  $ for the Kantowski-Sachs spacetime and for
$\omega=-1,$ that is $\Omega=1$. First row are plots with $\lambda=-1$ while
in the second row are plots with $\lambda=-5.$ The plots of the left and right
columns are the same but from a different view point. It is clear that there
are no attractors for the dynamical system and the trajectories start from the
finite regime, reach infinity and vice-versa. }%
\label{fig05}%
\end{figure}

In Figs. \ref{fig04} and \ref{fig05} we present three-dimensional phase-space
portraits for the field equations in the Poincare variables $\left(
\rho,\theta,\eta\right)  $ and different values of $\lambda$. We observe that
there is not any attractor for the dynamical system at the infinity, neither
in the finite regime as we derived in the previous Section for $\omega<0$.
Hence, the trajectories start from the finite regime, reach infinity and vice-versa.

\section{Dynamical analysis for the power-law potential}

\label{sec5}

Consider now the power-law potential $V\left(  \phi\right)  =V_{0}\phi^{\nu}$.
Then the dynamical equation (\ref{mm.10}) is%
\begin{equation}
\frac{d\lambda}{d\tau}=-\frac{2\sqrt{3}}{\nu}\lambda^{2}x. \label{mm.11}%
\end{equation}

We follow the same procedure as above and we determine the stationary points
$Q=\left(  \Sigma\left(  Q\right)  ,x\left(  Q\right)  ,y\left(  Q\right)
,\eta\left(  Q\right)  ,\lambda\right)  $ for the dynamical system
(\ref{mm.05})-(\ref{mm.08}) and (\ref{mm.11}) at the finite and infinity regimes

\subsection{Stationary points at the finite regime}

At the finite regime, the physically acceptable stationary points are
\[
Q_{1}^{\pm}=\left(  \pm\sqrt{1-\omega\left(  x_{1}\right)  ^{2}}%
,x_{1},0,1,0\right)  ,
\]
with the same existence conditions and physical properties as points
$P_{1}^{\pm}$.\qquad\
\[
Q_{2}^{\pm}=\left(  \pm\sqrt{1-\omega\left(  x_{2}\right)  ^{2}}%
,x_{2},0,-1,0\right)  ,
\]
with the same existence conditions and physical properties as points
$P_{2}^{\pm}$.%
\[
Q_{3}^{\pm}=\left(  \mp\sqrt{\frac{\omega}{\omega+48}},\mp\frac{4\sqrt{3}%
}{\sqrt{\omega\left(  \omega+48\right)  }},0,\pm2\sqrt{\frac{\omega}%
{\omega+48}}\right)  ,
\]
with the same existence conditions and physical properties as points
$P_{3}^{\pm}$.
\[
Q_{4}^{\pm}=\left(  0,0,1,\pm1,-4\right)  ,
\]
describes spatially flat FLRW geometry with $q\left(  Q_{4}^{\pm}\right)
=-1$, that is, the asymptotic solutions are de Sitter spacetimes

The main difference with the exponential potential above is the de Sitter
solutions. However, because the dimension of the dynamical system is greater
and a new dynamical variable exists the stability properties should be
investigated. Briefly, we discuss the stability properties for the new points
$Q_{4}^{\pm}$.

\subsubsection{Stability analysis}

For points $Q_{4}^{\pm}$ the eigenvalues of the linearized system are%
\[
e_{1}\left(  Q_{4}^{\pm}\right)  =-3~,~e_{2}\left(  Q_{4}^{\pm}\right)  =-2
\]%
\[
e_{3}\left(  Q_{4}^{\pm}\right)  =\mp\frac{3}{2}+\frac{1}{2}\sqrt
{\frac{3\left(  128+3\nu\omega\right)  }{\nu\omega}}~,~e_{4}\left(  Q_{4}%
^{\pm}\right)  =\mp\frac{3}{2}-\frac{1}{2}\sqrt{\frac{3\left(  128+3\nu
\omega\right)  }{\nu\omega}}%
\]
from which we conclude that the stationary point $Q_{4}^{-}$ is always a
saddle point, and the expanding de Sitter universe described by $Q_{4}^{+}$ is
an attractor for $\omega\nu>0$.

\begin{figure}[ptb]
\centering\includegraphics[width=1\textwidth]{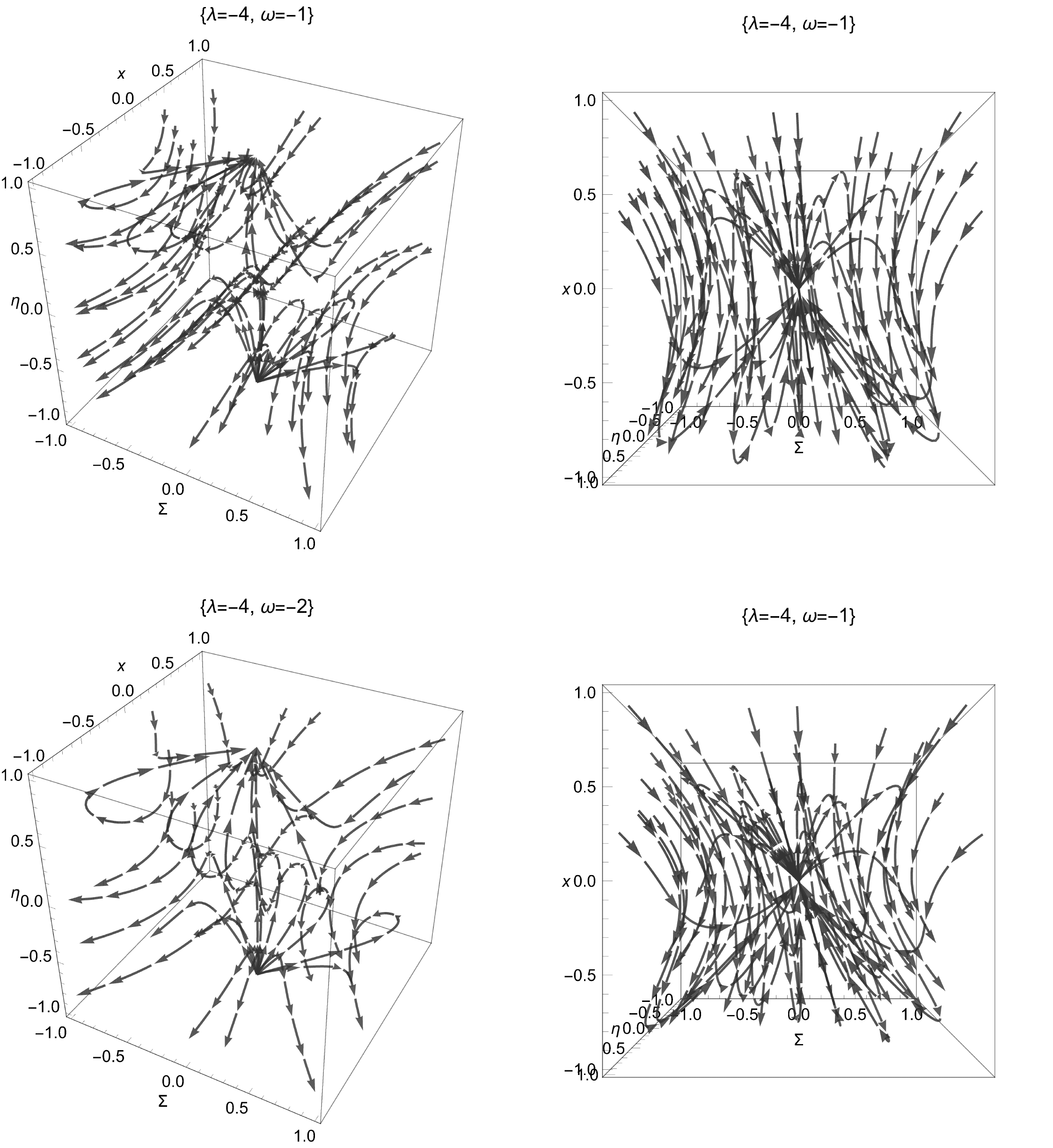}\caption{Phase-space
portraits for the dynamical system for the power-law potential in the
three-dimensional space $\left(  \Sigma,x,\eta\right)  .$ Plots are for
$\omega<0$ and $\omega\nu>0$. It is clear that the de Sitter point$~Q_{4}^{+}$
is an attractor at the finite regime. First row are plots for $\omega=-1$
while in the second row are plots with $\omega=-5.$ The plots of the left and
right columns are the same but from a different view point. }%
\label{fig06}%
\end{figure}

In Fig. \ref{fig06} we present the three-dimensional phase-portrait for the
field equations in which we observe that the de Sitter point $Q_{4}^{+}$ is an
attractor. The plots are for $\omega<0$ from which it is clear that the an
attractor exists at the finite regime.

\subsection{Stationary points at the infinity}

For $\omega<0$ the dynamical system can reach infinity. We consider the same
Poincare variables as before. We determine the same stationary points as
before where now $\lambda=0$.

The stability properties are the same hence we omit the presentation of the analysis.

We conclude that for the power-law potential the de Sitter solution can be an
attractor. In this case, the scalar-field potential acts as a cosmological
constant, which means that in the scalar-torsion theory the introduction of
the cosmological constant can solve the flatness and isotropic problem.

\section{Conclusions}

\label{sec6}

We investigated the phase-space for the gravitational field equations in
scalar-torsion theory with a Kantowski-Sachs background
spacetime.\ Specifically, we considered that the scalar field potential to be
the exponential potential, or the power-law potential, while we assumed that
there is not any other matter source. The scalar-torsion theory introduce a
constant parameter $\omega$ which indicate the coupling of the scalar field to
the teleparallel Lagrangian, in a similar way to the Brans-Dicke parameter, in
scalar-tensor theory and the Ricciscalar. 

For the Kantowski-Sachs spacetime we wrote the field equations in the
equivalent way of an algebraic-differential system, where we have used
dimensionless variables similar to that of the Hubble normalization. For the
latter dynamical system we derived the stationary points and we investigated
their stability properties. This analysis was necessary in order to
reconstruct the cosmological history. We remark that for $\omega>0$, the
phase-space is compact, however, when $\omega<0$ the dynamical variables can
reach the infinity regime.

For the exponential potential, and $\omega>0$ we found that there exist a
stationary point which can be an attractor, where the asymptotic solution
describes a Kantowski-Sachs universe which can be accelerated. Additionally,
there exist two set of points which describe unstable Kasner-like solutions.
Recall that a Kasner-like solution can describe the evolution of the physical
variables near to the cosmological singularity. Hence, for $\omega>0$ and the
exponential potential, the evolution of the Kantowski-Sachs universe is
simple. It can start from a singular solution and the dynamics to end into an
anisotropic Kantowski-Sachs space. On the other hand, for $\omega<0$ and the
exponential potential we found that there are not any attractors. Thus, the
trajectories of the dynamical system start from the finite regime, reach
infinity and vice-versa.

In the case of the power-law potential the behaviour of the dynamics is
different. In this case, the exist a stationary point which describe an
asymptotic isotropic, homogeneous and accelerated cosmological model, the de
Sitter spacetime. The de Sitter point exists in the de Sitter regime and it is
an attractor for $\omega>0$ and $\omega<0$. Finally, for $\omega<0$ the
analysis at the finite regime for the power-law potential is similar to the
analysis for the exponential potential.

We conclude that the scalar-tensor theory for an appropriate potential
function can solve the isotropization of the universe as also the zero valued
spatial curvature. In a future work we plan to investigated further the
evolution of anisotropies in scalar-tensor theory by studying the dynamics in
Bianchi geometries.

\textbf{Data Availability Statements:} Data sharing not applicable to this
article as no datasets were generated or analysed during the current study.

\begin{acknowledgments}
This work was partially financial supported by the National Research
Foundation of South Africa (Grant Numbers 131604). The author thanks for the
support of Vicerrector\'{\i}a de Investigaci\'{o}n y Desarrollo
Tecnol\'{o}gico (Vridt) at Universidad Cat\'{o}lica del Norte through
N\'{u}cleo de Investigaci\'{o}n Geometr\'{\i}a Diferencial y Aplicaciones,
Resoluci\'{o}n Vridt No - 096/2022.
\end{acknowledgments}

\end{document}